\documentstyle[12pt]{article}

\newcommand{\be}{\begin{equation}}
\newcommand{\bea}{\begin{eqnarray}}
\newcommand{\ba}{\begin{array}}
\newcommand{\bean}{\begin{eqnarray*}}
\newcommand{\ee}{\end{equation}}
\newcommand{\eea}{\end{eqnarray}}
\newcommand{\ea}{\end{array}}
\newcommand{\eean}{\end{eqnarray*}}

\def\D{\Delta}
\def\hPi{\widehat{\Pi}}
\def\hGa{\widehat{\Gamma}}
\def\hSi{\widehat{\Sigma}}
\def\Sm{$S$-matrix~}
\def \dsl {\partial \kern-.55em{/}}
\def \Dsl {D \kern-.65em{/}}
\def \qsl {q \kern-.45em{/}}
\def \slp {p \kern-.45em{/}}
\def \ksl {k \kern-.45em{/}}

\def \Gfs {Green's functions~}

\newcommand{\PRD}{ {\it Phys. Rev.}~{\bf D}}
\newcommand{\PLB}{ {\it Phys. Lett.}~{\bf B}}
\newcommand{\NPB}{ {\it Nucl. Phys.}~{\bf B}}

\begin{document}

\begin{titlepage}
\rightline{ NYU-TH-96/6/01}
\rightline{June 1996}

\vspace*{1.5cm}

\begin{center}
               {\Large\bf The dual gauge fixing property of the \Sm }
\end{center}

\vspace*{1cm}
\begin{center}
{\bf Joannis Papavassiliou}  
\end{center}
\begin{center}
{\it Centre de Physique Th\'eorique, CNRS, Luminy,

 F-13288 Marseille, France}

\end{center}
\vspace*{0.2cm}
\begin{center}
{\bf and}
\end{center}
\vspace*{0.1cm}

\begin{center}
{\bf Kostas Philippides}
\end{center}
\begin{center}
           {\it New York University, Department of Physics

      Andre and Bella Meyer Hall of Physics, 4 Washington Place,

                    New York, NY 10003, USA}
\end{center}

\vspace*{1cm}
\begin{center}

{\bf ABSTRACT}
\end{center}
\vspace*{0.5cm}
\baselineskip=25pt

The $S$-matrix is known to be independent of the gauge fixing parameter to 
all orders in perturbation theory.
In this paper by employing the pinch technique we prove at one loop a 
stronger version of this independence.
In particular we show that one can use a gauge fixing parameter for the 
gauge bosons inside quantum loops
which is different from that used for the bosons outside loops, and the 
$S$-matrix is independent from 
both. Possible phenomenological applications of this result are briefly 
discussed. 

\end{titlepage}

\baselineskip=25pt

\setcounter{equation}{0}
\section{Introduction}

In this paper we will discuss an interesting  
property of the $S$-matrix of gauge theories, 
which is easy to prove for QED, but  
is not at all evident for non-abelian theories 
such as QCD, or the electroweak $SU(2)_L\times U(1)_Y$ model.

As a result of the quantization of a gauge theory, 
arbitrary gauge fixing parameters (GFP), which we will 
collectively denote by $\xi$, infest the  
Feynman rules used in perturbative calculations. 
It is well known however 
that even though individual Feynman diagrams are GFP-dependent,
when combined to form the $S$-matrix element of a physical process, 
they give rise to GFP-independent expressions, order by order 
in perturbation theory
\cite{GIS}.
It turns out that a stronger version of this GFP cancellation exists, 
which we will prove at one loop order.

We will separate the virtual gauge bosons of a Feynman graph 
into two classes: the ``loop'' gauge bosons, i.e. those virtual gauge bosons
 which appear inside the loops of a Feynman graph, and the ``tree'' 
gauge bosons, 
which are not part of 
a loop. In other words, the ``loop'' gauge bosons 
are irrigated by the virtual
loop momentum we integrate over, while the ``tree'' gluons are not.
The tree-level propagators of each gauge bosons in either class depend on 
$\xi$.

We will now go one step further and make the arbitrary replacements 
$\xi \rightarrow \xi_t$ for the propagators of 
the tree gauge bosons and $\xi \rightarrow \xi_l$
for the propagators of the loop gauge bosons, where 
$\xi_t \not =  \xi_l$.
In this way one introduces in general two entirely different 
gauge-fixing parameters ($\xi_t$ and $\xi_l$).
Then one can show that 
the $S$-matrix is unchanged, and that it
 is invariant under the {\it separate} change :
\bea 
(D) ~: & ~\xi_t \rightarrow \xi_t^{'} \nonumber \\
      & ~ \xi_l\rightarrow\xi_l' 
\eea
or equivalently, that it is
independent of {\it  both } $\xi_t$ and $\xi_l$.
In particular, one can prove the above statement
{\it before} any momentum-integration is carried out.

To clarify the previous procedure, we consider a particular example. 
We start with the usual classical QCD Lagrangian density
\be
{\cal L}_C= -\frac{1}{4}{\cal F}_{\mu\nu}^{a}{\cal F}^{a\mu\nu}
          + \bar {\psi}(i\Dsl-m){\psi} ~, 
\ee
which is invariant under the gauge transformations 
\bea
A^{'a}_{\mu}(x) &=& U(x)A^{a}_{\mu}(x)U^{-1}(x)- 
[\partial_{\mu}U(x)]U^{-1}(x)~, 
\\
\psi^{'}(x)&=& U(x)\psi (x)~, 
\\
U(x)&=&\exp (-i\omega^{a}(x)T^a)~, 
\eea
where $T^a$ are the matrix representations of the SU(3) group. 
We then quantize ${\cal L}_C$ using the gauge fixing term
${\cal L}_{GF}=-\frac{1}{2\xi}(G^{a})^2 =  
-\frac{1}{2\xi}{(\partial^{\mu} A^a_{\mu})}^{2}$, 
and the corresponding Fadeev-Popov term 
${\cal L}_{\Phi\Pi}=\bar{c^{a}}\frac{\delta G^{a}}{\delta \omega^{b}} c^{b}
=\bar{c^{a}}(-\partial^{\mu}D_{\mu}^{ab})c^{b}$,   
\be
{\cal L}_Q= {\cal L}_C+{\cal L}_{GF}+{\cal L}_{\Phi\Pi} ~.
\ee
Then use the Feynman rules obtained from the ${\cal L}_Q$
Lagrangian density to compute the one-loop $S$-matrix
element $T$, for elastic scattering of quarks 
$q_{1}q_{2}\rightarrow q_{1}q_{2}$,
with masses $m_{1}$ and  $m_{2}$. 
In particular, the gluon propagator reads
\be
i\Delta_{\mu\nu}(q,\xi) =  \frac{-i}{q^2}
[g_{\mu\nu}-(1-\xi)\frac{q_{\mu}q_{\nu}}{q^2}]~,
\label{GaugeProp}
\ee
and the ghost propagator
\be
i\Delta_{c}= \frac{i}{q^2} ~.
\label{GhostVProp}
\ee
Let us call the integration 
momentum $k$.
Self-energy, vertex and box graphs will 
contribute respectively to the $T_1,~T_2$ and $T_3$ parts of the amplitude, i.e.
\bea 
T(s,t,m_1,m_2) &= &
T_1(t,\xi)+T_2^{(1)}(t,m_1,\xi)+T_2^{(2)}(t,m_2,\xi)
\nonumber \\
&& +T_3(s,t,m_1,m_2,\xi) ~,
\eea
where $t= -q^2$, $q=p_1-p'_1=p'_2-p_2$ and $p_i,~p_i'$ are respectively the 
initial and final momenta of the quarks. 
Label by $\Delta_{\mu\nu}^{t}(q,\xi)$ the propagators of the tree gluons, and
by $\Delta_{\mu\nu}^{l}(k,\xi)$ the propagators of the loop gluons 
Then replace $\Delta_{\mu\nu}^{t}(q,\xi)\rightarrow 
\Delta_{\mu\nu}^{t}(q,\xi_{t})$ and 
$\Delta_{\mu\nu}^{l}(k,\xi)\rightarrow\Delta_{\mu\nu}^{l}(q,\xi_{l})$,
where $\xi_{t}\neq\xi_{l}$, in general. 
The above
transformation does not change the value of the $S$-matrix
element, i.e. $S$ is independent of {\it both} $\xi_{t}$ and $\xi_{l}$.

Of course, instead of the $R_{\xi}$ gauges,
one could choose
a different gauge-fixing scheme.
In the case
of a ghost-free non-covariant gauge
such as the light-cone gauge \cite{axial}
for example,
the gauge fixing term is ${\cal L}_{LC}=
-\frac{1}{2\xi}{(n_{\mu}A^{\mu})}^{2}$,
where $n_{\mu}$ is an arbitrary four-vector, for which 
$n_{\mu}A^{\mu}=0$ and $n^{2}=0$; 
the corresponding tree-level gluon propagator in the $\xi \rightarrow 0$ 
limit is 
given by 
\be
i\Delta_{\mu\nu}(q,n) = \frac{-i}{q^2}\left[g_{\mu\nu}-
\frac{n_{\mu}q_{\nu}+n_{\nu}q_{\mu}}{n\cdot q} \right] ~.
\label{LightCone}
\ee
Carrying out the corresponding replacement 
$\Delta_{\mu\nu}^{t}(q,n)\rightarrow \Delta_{\mu\nu}^{t}(q,n_{t})$
and $\Delta_{\mu\nu}^{l}(q,n)\rightarrow \Delta_{\mu\nu}^{l}(q,n_{l})$  
we will find that the S-matrix is independent of both
$n_{t}$ and $n_{l}$. We call the above property, the ``dual'' gauge-fixing 
(DGF) property of the $S$-matrix. 

The paper is organized as follows: In section 2 we prove the DGF property 
for the case of QED, and 
discuss the basic ingredients which are crucial for the proof. In section 3 
we extend the proof
to the case of non-Abelian gauge theories; for the case of QCD we show that 
the features which operate
in the QED case 
 are concealed by the conventional perturbative formulation, but they can 
be exposed by resorting to
the systematic rearrangement of graphs dictated by the pinch technique (PT)
~\cite{Cornwall}, \cite{PTvarious}.
In section 4 we extend this analysis to the electroweak sector, where exactly 
analogous results apply.
Finally, in section 5, we discuss our conclusions, and briefly present some 
possible applications of
the DGF property in the context of the electroweak phenomenology. 

\setcounter{equation}{0}
\section {The QED case}

In order 
to understand the mechanism which enforces the DGF property at the level of 
the $S$-matrix, 
let us focus for
a moment on QED. 
In QED the above property can easily be proved;
this is so because the photon self-energy 
$\Pi_{\mu\nu}(q)$, the photon-electron vertex $\Gamma_{\mu}$,
and the electron self-energy $\Sigma(p)$ have the following
properties (at least at one loop).

(a) $\Pi_{\mu\nu}(q)$ is GFP-independent,
and transverse, i.e. $q^{\mu}\Pi_{\mu\nu}(q)=0$.

(b) $q^{\mu}\Gamma_{\mu}(p_{1},p_{2})=e[\Sigma(p_{1})-\Sigma(p_{2})]$, 
by virtue of QED 
the Ward identity.

(c) the sum of the two box diagrams (direct and crossed) 
is GFP-independent.

From (a), (b), and (c) follows that the improper vertex 
$G_{\mu}$, which consists of $ \Gamma_{\mu}(p_{1},p_{2})$
and the wave function corrections to the external fermion legs, is  
GFP-independent, UV finite, and transverse, i.e.
$q^{\mu}G_{\mu}=0$ .

It is now easy to see how the DGF property of the S-matrix
holds in the case of QED.
To begin with, the box diagrams contain only loop photons
$\Delta_{\mu\nu}^{i}$, and their sum is GFP-independent, so it is invariant 
under the transformation (D): $T\stackrel{(D)}{\rightarrow} T'$ , 
namely $T_{3}\equiv T'_{3}$. 
The photon self-energy $\Pi_{\mu\nu}$ at one loop consists of a fermion loop,
so its value does not change under (D). In addition, 
any gauge fixing parameters stemming from $\Delta_{\mu\nu}^{t}$
vanishes, because it either gets contracted with the external conserved 
current,
or with the transverse $\Pi_{\mu\nu}$.
So the part $T_{1}$ 
of the $S$-matrix 
before the
transformation reads
\bea
T_{1}&=&{\bar{u}}_{1}\gamma_{\rho}u_{1}\Delta^{\rho\mu}(q,\xi)
\Pi_{\mu\nu}(q)\Delta^{\nu\sigma}(q,\xi)
{\bar{u}}_{2}\gamma_{\sigma}u_{2}\nonumber\\
&=& {\bar{u}}_{1}\gamma_{\mu}u_{1}[\frac{1}{q^{2}}]
\Pi^{\mu\nu}(q)[\frac{1}{q^{2}}]{\bar{u}}_{2}\gamma_{\nu}u_{2}~.
\label{T1QED}
\eea
After the transformation (D), $T_{1}\stackrel{(D)}{\rightarrow} T_{1}^{'}$,
with $T_{1}^{'}$ given by
\bea
T_{1}^{'}&=&{\bar{u}}_{1}\gamma_{\rho}u_{1}\Delta^{\rho\mu}(q,\xi_t)
\Pi_{\mu\nu}(q)\Delta^{\nu\sigma}(q,\xi_t)
{\bar{u}}_{2}\gamma_{\sigma}u_{2}\nonumber \\
&=&{\bar{u}}_{1}\gamma_{\mu}u_{1}[\frac{1}{q^{2}}]
\Pi^{\mu\nu}(q)[\frac{1}{q^{2}}]{\bar{u}}_{2}\gamma_{\nu}u_{2}\nonumber \\
&=&T_{1} ~.
\label{T12QED}
\eea
Notice that in the $R_{\xi}$ gauges, due to current conservation, the 
dependence on
$\xi$ vanishes, even without using the transversality of 
$\Pi_{\mu\nu}(q)$.
If instead we had been working in a non-covariant gauge, 
 we would have to  use the WI 
$q^{\mu}\Pi_{\mu\nu}(q)=0$ in the above equation,
 because the terms proportional to  
$n_t^{\rho}q^{\mu}$ and $n_t^{\sigma}q^{\nu}$ 
cannot be contracted with the external conserved current.

Finally, for the part $T_{2}$ of the $S$-matrix involving the
improper vertex $G_{\mu}$ we have at the beginning:
\bea
T_{2}&=&{\bar{u}}_{1}\gamma_{\rho}u_{1}\Delta^{\rho\mu}(q,\xi)
{\bar{u}}_{2}G_{\mu}u_{2}\nonumber \\
&=&{\bar{u}}_{1}\gamma_{\mu}u_{1}[\frac{1}{q^{2}}]{\bar{u}}_{2}
G^{\mu}u_{2} ~.
\label{T2QED}
\eea
On the other hand, after imposing (D):
\bea
T_{2}^{'}&=& {\bar{u}}_{1}\gamma_{\rho}u_{1}\Delta^{\rho\mu}(q,\xi_t)
{\bar{u}}_{2}G_{\mu}u_{2}\nonumber \\
&=& T_{2} ~.
\label{T22QED}
\eea
Again, if we were to work in a non-covariant gauge 
we would need to resort to the transversality of $G_{\mu}$,
i.e. use that $q^{\mu}G_{\mu}=0$.

Finally, since $T_{i}^{'}= T_{i}$, for $i=1,2,3$, the 
$S$-matrix is invariant under (D).

Even though the above proof is very straightforward, it allows  
one to recognize the crucial ingredients which enforce the invariance 
under (D). They are :

(a) The fact that certain Green's functions are GFP-independent 
in any gauge-fixing procedure.

(b) The fact that in QED the Green's functions 
satisfy their naive, tree-level Ward identities,
even {\it after} quantum corrections have been taken into account.

\setcounter{equation}{0}
\section{Non-Abelian gauge theories: The QCD case}

The previous 
proof of the DGF property, 
which is very transparent in the case of QED, becomes 
complicated in the case of non-Abelian gauge theories (NAGT), 
such as QCD, or theories with Higgs mechanism such as the
$SU(2)_{L}\times U(1)_{Y}$ electroweak sector 
of the standard model. 
The reason is that in the conventional formulation of NAGT
the two crucial properties mentioned above fail to be satisfied.
Regarding property (a), in NAGT the gauge boson self-energy is GFP-{\it dependent}, 
already at one loop \cite{FLS}, \cite{GW}. As for property (b), 
after quantization the tree level Ward identities are replaced
by complicated Slavnov-Taylor identities, derived from the residual 
BRST symmetries.
However, as we will explicitly illustrate, 
the DGF property holds {\it also} 
for these theories, at least at one-loop.
 
Let us first concentrate on a QCD example and examine 
at one-loop the $S$-matrix element 
for quark-antiquark annihilation into a pair of gluons ($g$), i.e.
the process
$q(p_1)\bar{q}(p_2) \rightarrow g(q_1) g(q_2)$. 
This process contains both the  $gq\bar{q}$ vertex as well as the 
three gluon vertex at one loop. The \Sm element is again 
decomposed into self-energy, vertex, and box parts, 
\bea
T(s,t,m) &=&
T_1(q,\xi)+T_2^{f}(p_1,p_2,m,\xi)+
T_2^{g}(q_1,q_2,\xi)
\nonumber \\
&& +T_3(s,t,m,\xi) ~,
\eea
where the superscript ``$f$''(``$g$'') in $T_{2}$ refers to the two external 
 ``on-shell'' fermions (gluons).
Under the transformation (D), the sub-amplitudes assume
the following forms: 

The self energy sub-amplitude is: 
\be 
T_1(s,\xi_t,\xi_l) = \bar{u}_1\gamma_{\alpha}u_2 \Delta^{\alpha\mu}(q,\xi_t)
\Pi_{\mu\nu}(q,\xi_l)\Delta^{\nu\beta}(q,\xi_t)
\Gamma^{(0)}_{\beta\rho\sigma}(q,q_1,q_2) \epsilon_{1}^{\rho}\epsilon_{2}^{\sigma}
\ee
where 
$\Gamma^{(0)}_{\beta\rho\sigma}(q,q_1,q_2)$ is the usual tree-level three-gluon vertex
\be
\Gamma^{(0)}_{\beta\rho\sigma}(q,q_1,q_2)= {(q-q_{1})}_{\sigma}g_{\beta\rho}
+ {(q_{1}-q_{2})}_{\beta}g_{\rho\sigma} + {(q_{2}-q)}_{\rho}g_{\sigma\beta} ~,
\ee
and $\epsilon_{i}^{\mu}$, $i=1,2$ are the polarization vectors 
corresponding to the external gluon with momentum $q_{i}$; clearly,
$q_{i}\cdot\epsilon_{i}=0$. 
The vertex parts, together with the external leg corrections, 
are:
\bea
T_2^{f}(s,m,\xi_t,\xi_l) =  
& \bar{u}_1\Gamma_{\alpha}^{(1)}(p_1,p_2,q;\xi_l)u_2
\Delta^{\alpha\mu}(q,\xi_t)
\Gamma^{(0)}_{\mu\rho\sigma}(q,q_1,q_2) \epsilon_{1}^{\rho}\epsilon_{2}^{\sigma}
\nonumber \\
& + \bar{u}_1\Sigma(p_1;\xi_l)\frac{1}{\slp_1-m}\gamma_{\alpha}u_2 
\Delta^{\alpha\mu}(q,\xi_t)
\Gamma^{(0)}_{\mu\rho\sigma}(q,q_1,q_2) \epsilon_{1}^{\rho}\epsilon_{2}^{\sigma}
\nonumber \\
& + \bar{u}_1\gamma_{\alpha}\frac{1}{\slp_2-m}\Sigma(p_2;\xi_l)u_2
\Delta^{\alpha\mu}(q,\xi_t)
\Gamma^{(0)}_{\mu\rho\sigma}(q,q_1,q_2) \epsilon_{1}^{\rho}\epsilon_{2}^{\sigma}~,
\eea
\bea
T_2^{g}(s,\xi_t,\xi_l)&=&
\bar{u}_1\gamma_{\alpha}u_2 \Delta^{\alpha\mu}(q,\xi_t)
\Gamma_{\mu\rho\sigma}^{(1)}(q,q_1,q_2;\xi_l)\epsilon_{1}^{\rho}\epsilon_{2}^{\sigma}
\nonumber \\
&& + \bar{u}_1\gamma_{\alpha}u_2 \Delta^{\alpha\mu}(q,\xi_t)
\Gamma^{(0)}_{\mu\beta\sigma}(q,q_1,q_2)\Delta^{\beta\nu}(q_1,\xi_t) 
\Pi_{\nu\rho}(q_1;\xi_l) \epsilon_{1}^{\rho}\epsilon_{2}^{\sigma}
\nonumber \\
&& + \bar{u}_1\gamma_{\alpha}u_2 \Delta^{\alpha\mu}(q,\xi_t)
\Gamma^{(0)}_{\mu\rho\beta}(q,q_1,q_2) \Delta^{\beta\nu}(q_2,\xi_t) 
\Pi_{\nu\sigma}(q_2;\xi_l)\epsilon_{1}^{\rho}\epsilon_{2}^{\sigma}~,
\eea
Finally, the box is given by: 
\be
T_3(s,t,m,\xi_l)  = B(p_1,p_2,q_1,q_2,m,\xi_l)_{\rho\sigma}
\epsilon_{1}^{\rho}\epsilon_{2}^{\sigma}~.
\ee

To prove that the \Sm element is independent of both $\xi_t$ and $\xi_l$ 
we proceed as follows: 

The first step is to show that the dependence on $\xi_l$ cancels 
regardless of what one chooses for $\xi_t$. To this end we employ the 
PT. The PT rearranges 
the Feynman diagrams by appropriately exploiting the following two 
elementary Ward identities, satisfied by the tree level 
$gf\bar{f}$ and $ggg$ vertices respectively: 

\be
k_{\mu}\gamma^{\mu} \equiv \ksl = (\ksl+\slp-m) - (\slp-m) ~,
\label{WI1}
\ee
\be
k^{\mu}{\Gamma}_{\mu\nu\alpha}^{(0)}(k,p-k,p)=
  (p-k)^2 t_{\nu\alpha}(p-k)  -  p^2 t_{\nu\alpha}(p) ~, 
\label{WI2}
\ee
where $t_{\alpha \beta}(q) = g_{\alpha \beta}-q_{\alpha}q_{\beta}/q^2$ is the 
usual transverse projector. 

Before carrying out any calculations, we first let the longitudinal
momenta supplied by the 
gluon propagators 
or the trilinear gluon vertices trigger the above WI.
The inverse propagators thus generated 
will either vanish on shell or cancel (pinch) 
an internal fermionic or bosonic propagator inside the loop.
As a result of these cancellations, 
 parts from the 
vertex or box graphs will emerge,  
which will have the same kinematic structure as the self-energy graphs.
The final step of casting these  expressions into the 
desired form of the  self-energy graphs as in $T_1$, is to recognize 
that a tree-level gluon propagator must be attached at the point 
where pinching took place.  For this purpose unity is inserted 
in the form of a propagator times its inverse, using the following 
elementary identity which holds for any gauge fixing procedure 
(covariant, non-covariant, etc.)
\bea
   g_{\alpha}^{\beta}&=&\Delta_{\alpha \mu}(q;\xi_t)
[\Delta^{-1}]^{\mu \beta}(q;\xi_t)
   =\Delta_{\alpha \mu}(q;\xi_t)[-q^2t^{\mu \beta}] + ... 
       \nonumber \\
           &=&\Delta^{-1}_{\alpha \mu }(q;\xi_t)\Delta^{\mu \beta}(q;\xi_t) =
        [-q^2 t_{\alpha \mu}]\Delta^{\mu \beta}(q;\xi_t) + ... 
\label{Identity}
\eea
where the ellipses denote terms that will 
vanish when 
contracted either with $\bar{u}_1\gamma_{\alpha}u_2$ or 
$\Gamma^{(0)}_{\beta\rho\sigma}(q,q_1,q_2) \epsilon_{1}^{\rho}\epsilon_{2}^{\sigma}$.
The $q^2 t_{\alpha \mu}$ factor will be part of the pinch expression and 
it is manifestly gauge independent.
It is important to emphasize that no $\xi_l$ dependences have been introduced 
in this step. 
Subsequently, the pinch parts extracted from the vertex and box graphs 
are alloted to the usual self-energy graphs, in order 
to define a new effective one-loop self-energy for the gluon. 
As has been shown 
by  explicit calculations in a wide variety of gauges 
\cite{Cornwall},\cite{PaBFM},\cite{Sasaki} 
(non-covariant, covariant, background),
and recently by rigorous arguments based on analyticity, unitarity, and BRST symmetry
\cite{PP2}
 this rearrangement suffices 
to cancel all dependence on $\xi_{l}$ inside the loop integrals. 
The crucial point is that the $\xi_{l}$- cancellations takes place in a 
kinematically distinct way, i.e. one ends up with propagator, vertex, 
and box-like
structures, which are {\it individually} independent of $\xi_{l}$. 
Thus after the PT rearrangement the sub-amplitudes assume the form:
\be 
T_1(q,\xi_t) = \bar{u}_1\gamma_{\alpha}u_2 \Delta^{\alpha\mu}(q,\xi_t)
\hPi_{\mu\nu}(q)\Delta^{\nu\beta}(q,\xi_t)
\Gamma^{(0)}_{\beta\rho\sigma}(q,q_1,q_2) \epsilon_{1\rho}\epsilon_{2\sigma} ~,
\ee
\bea
T_2^{f}(p_1,p_2,m,\xi_t) &=&  \bar{u}_1\hGa_{\alpha}^{(1)}(p_1,p_2,q)u_2
\Delta^{\alpha\mu}(q,\xi_t)
\Gamma^{(0)}_{\mu\rho\sigma}(q,q_1,q_2) \epsilon_{1}^{\rho}\epsilon_{2}^{\sigma}
\nonumber \\
&& + \bar{u}_1\hSi(p_1)\frac{1}{\slp_1-m}\gamma_{\alpha}u_2 
\Delta^{\alpha\mu}(q,\xi_t)
\Gamma^{(0)}_{\mu\rho\sigma}(q,q_1,q_2) \epsilon_{1}^{\rho}\epsilon_{2}^{\sigma}
\nonumber \\
&& + \bar{u}_1\gamma_{\alpha}\frac{1}{\slp_2-m}\hSi(p_2)u_2
\Delta^{\alpha\mu}(q,\xi_t)
\Gamma^{(0)}_{\mu\rho\sigma}(q,q_1,q_2) \epsilon_{1}^{\rho}\epsilon_{2}^{\sigma}~, 
\eea
the vertex parts together with the corrections for the external legs 
are
\bea
T_2^{g}(q_1,q_2,\xi_t,\xi_l)&=&
\bar{u}_1\gamma_{\alpha}u_2 \Delta^{\alpha\mu}(q,\xi_t)
\hGa_{\mu\rho\sigma}^{(1)}(q,q_1,q_2)\epsilon_{1}^{\rho}\epsilon_{2}^{\sigma}
\nonumber \\
&& + \bar{u}_1\gamma_{\alpha}u_2 \Delta^{\alpha\mu}(q,\xi_t)
\Gamma^{(0)}_{\mu\beta\sigma}(q,q_1,q_2)\Delta^{\beta\nu}(q_1,\xi_t) 
\hPi_{\nu\rho}(q_1) \epsilon_{1}^{\rho}\epsilon_{2}^{\sigma}
\nonumber \\
&& + \bar{u}_1\gamma_{\alpha}u_2 \Delta^{\alpha\mu}(q,\xi_t)
\Gamma^{(0)}_{\mu\rho\beta}(q,q_1,q_2) \Delta^{\beta\nu}(q_2,\xi_t) 
\hPi_{\nu\sigma}(q_2)\epsilon_{1}^{\rho}\epsilon_{2}^{\sigma} ~,
\eea
and finally the box-like contributions
\be
T_3(p_1,p_2,q_1,q_2,m)  = \widehat{B}(p_1,p_2,q_1,q_2,m)_{\rho\sigma}
\epsilon_{1}^{\rho}\epsilon_{2}^{\sigma}~.
\ee
The hatted quantities in the above expressions denote the PT effective Green's
functions, which are manifestly independent of $\xi_{l}$; their 
exact closed expressions have been reported elsewhere
\cite{PTvarious}, and  are not important for the subsequent analysis.

The second step in the proof is to observe that the new effective one-loop 
\Gfs constructed via the PT in the first step satisfy their respective 
{\it tree-level} WI. It is important to emphasize that these 
classical WI are now valid even after the one-loop quantum corrections 
have been taken into account. This is to be contrasted to the complicated 
Slavnov-Taylor identities that the one-loop \Gfs usually satisfy. 
One can easily verify for  the PT \Gfs that~:
\be
q^{\mu}\hPi_{\mu\nu}(q) = 0 ~,
\ee
\be
q^{\mu}\hGa_{\mu}(q,p_1,p_2) = g\left[\hSi (p_1)- \hSi(p_2)\right]~,
\ee
\be
q^{\mu}\hGa_{\mu\rho\sigma}(q,q_1,q_2) =g\left[ \hPi_{\rho\sigma}(q_{1}) 
-\hPi_{\rho\sigma}(q_{2})\right]~,
\ee
where $g$ is the gauge coupling. 
Consequently, the improper vertices $\widehat{G}_{\mu}$ and 
$\widehat{G}_{\mu\rho\sigma}$ which contain the corrections to the 
external fermion or gluon legs are transverse; 
$q^{\mu}\widehat{G}_{\mu}=0$, $q^{\mu}\widehat{G}_{\mu\rho\sigma}=0$. 
Using the above property, it is now straightforward to show that the residual $\xi_t$ 
dependence cancels within each sub-amplitude, and that the \Sm element is 
independent of $\xi_t$. At this point it is important to note that 
the key element to the proof has been the PT rearrangement, which 
transforms the ordinary sub-amplitudes $T_i$ to hatted ones, $\widehat{T}_i$,  
{\it without} mixing the ``loop'' GFP $\xi_l$ with the ``tree'' GFP $\xi_t$. 
Exactly as in QED, the new sub-amplitudes consist of one-loop 
\Gfs which are independent 
of $\xi_l$, and  satisfy their tree level WI; this last property in 
turn eliminates all 
remaining $\xi_t$ dependences.

\setcounter{equation}{0}
\section {The Electroweak sector}

The previous arguments can be generalized
to the case of a non-Abelian 
theory with tree-level symmetry breaking, such as the
 $SU(2)_L\times U(1)_Y$ electroweak model.
 Even though the equivalent proof is technically more involved, mainly because
in the electroweak sector the currents are not conserved, 
and the presence of additional unphysical degrees of freedom
(such as the would-be Goldstone bosons) complicates matters considerably, 
the conceptual issues remain the same. 
One needs to construct effective Green's functions which
are manifestly GFP-independent, and, in addition, they satisfy tree-level
Ward identities, even at one loop. 
Both of these requirements can be satisfied when one resorts to the PT 
rearrangement of the $S$-matrix \cite{EW}. 

Let us  concentrate on the $S-$matrix element of a charged four-fermion 
process, and work in the renormalizable $R_{\xi}$ class of gauges.  
We consider the scattering $i_u i_{\bar{d}} \rightarrow f_u f_{\bar{d}}$, 
where $i$ and $f$ are the
initial and final $SU(2)$ fermion doublets respectively, with masses 
$m_{\{i\}}=m_u,m_d$ and  $m_{\{f\}}=M_u,M_d$, and momenta 
$p_u,p_d$ and $l_u,l_d$, where $q=p_u-p_d= l_d-l_u$.
The $S-$matrix element consists again of the sub-amplitudes 
$T_1(s;\xi_{j}),~T^i_2(s,m_{\{i\}};\xi^{j})$,  
$T^f_2(s,m_{\{f\}};\xi^{j})$ and 
$T_3(s,t,m_{\{i\}},m_{\{f\}};\xi^{j})$; they depend explicitly 
on the gauge fixing parameters $ \xi^{j}$, where $j=W,Z,\gamma$. 
We now show that by replacing $\xi \rightarrow \xi_{t}$ outside of the loops
(there is only one gauge parameter outside the loops, namely $\xi=\xi_{W}$) 
and $\xi^{j} \rightarrow \xi_{l}^{j}$ inside the loops, for all $ \xi_{j}$,
 the 
$S-$matrix that consists of the sum $T_1+T_2+T_3$ remains unchanged.
After the above replacement the amplitudes read :
\bea
T_1(t;\xi_{t}, \xi_{l}^{j}) &=&J_{W\alpha} \D_W^{\alpha\mu}(q, \xi_{t})
\Pi ^W_{\mu\nu}(q,\xi_{l}^{j})\D_W^{\nu\beta}(q, \xi_{t}) J^{+}_{W\beta}
\nonumber \\ 
&&+J_{\phi}\D_{\phi}(q, \xi_{t})\Pi^-_{\nu}(q,\xi_{l}^{j})
\D_W^{\nu\beta}(q, \xi_{t}) J^{+}_{W\beta} \nonumber \\ 
&& +J_{W\alpha} \D_W^{\alpha\mu}(q, \xi_{t})\Pi ^-_{\mu}(q,\xi_{l}^{j})
\D_{\phi}(q, \xi_{t})J^{+}_{\phi}\nonumber \\ 
&& + J_{\phi}\D_{\phi}(q, \xi_{t}) \Pi ^{\phi}(q,\xi_{l}^{j})
 \D_{\phi}(q, \xi_{t})J^{+}_{\phi} 
\label{T1SM}
\eea
\bea
T_2(t,m_{\{i\}},m_{\{f\}};\xi_{t}, \xi_{l}^{j})&=&
\Gamma^{W^{-}i_u\bar{i}_d} _{\alpha}(-q,p_u,-p_d;\xi_{l}^{j})
\D_W^{\alpha\beta}(q, \xi_{t}) J^{+}_{W\beta} \nonumber \\ 
&&+\Gamma^{\phi^{-}i_u\bar{i}_d}(-q,p_u,-p_d;\xi_{l}^{j})
\D_{\phi}(q, \xi_{t})J^{+}_{\phi} \nonumber \\ 
&&+ J_{W\alpha}\D_W^{\alpha\beta}(q,\xi_t)
\Gamma^{W^{+}\bar{f}_u f_d}_{\beta}(q,-l_u,l_d;\xi_{l}^{j}) \nonumber \\ 
&&+J_{\phi}\D_{\phi}(q, \xi_{t})
\Gamma^{\phi^{+}\bar{f}_u f_d}(q,-l_u,l_d;\xi_{l}^{j}) \\
&&+ ~external ~leg~corrections 
\label{T2SM}
\eea
\be
T_3(s,t,m_u,m_d,M_u,M_d;\xi_{l}^{j})~ \equiv ~
B(s,t,m_u,m_d,M_u,M_d;\xi_{l}^{j}) \hskip1.5cm
\label{T3SM}
\ee

We now use the PT to rearrange the above amplitudes by  employing the 
tree level Ward identity of the vertex $Wf\bar{f'}$ 
\be
k_{\mu}\gamma^{\mu} P_{L}\equiv \not k  P_{L} = 
S_{i}^{-1}(t+k) P_{L} - P_{R}S_{j}^{-1}(t) + m_{i}P_{L} - m_{j}P_{R}
\label{BrokenPinch}
\ee
where $k_{\mu}$ is a loop integration momentum and $t=p,l$ is 
one of the external momenta.
As in the QCD case, the action of the first term in 
Eq.(\ref{BrokenPinch})  
is to cancel the fermion propagator of the loop, while the second 
vanishes on shell. 
Such $k_{\mu}$  momenta are 
provided inside the loops by the three gauge boson vertices, 
the longitudinal parts
of the gauge boson propagators, and by the gauge-scalar-scalar vertices. 
This procedure allows us to extract from the 
box amplitude $T_3$ pieces, which exhibit either the propagator-like structure of 
$T_1$ or the vertex-like structure of $T_2$, depending on how many internal 
fermion propagators have been cancelled \cite{Comment}. 
Similarly from the vertex 
amplitude $T_2$ we extract the parts that have the propagator-like 
structure of $T_1$. These pinch parts are appended to the relevant amplitudes 
and define the new $\hat{T}_1,~\hat{T}_2$ and $\hat{T}_3$; they can be obtained
from the expressions of Eq.(\ref{T1SM}) - Eq.(\ref{T3SM}) by substituting 
$\Pi \rightarrow \hPi,~\Gamma \rightarrow \hGa,~ B \rightarrow \widehat{B}$. 
Again, as has been verified by explicit calculations \cite{EW}, all 
$\xi_l$ dependence in the above amplitudes has cancelled,  
and this has happened completely independently of what $\xi_t$ is. 

We now show that the $\xi_t$ dependence also cancels in these amplitudes. 
This final cancellation is enforced by a set of Ward identities that the 
new hatted, manifestly  $\xi_l$ independent, Green's functions satisfy. 
The PT self energy functions have been shown to 
satisfy the following WI ~\cite{EW}
\bea
q^{\mu}\hPi^W_{\mu\nu}(q) \mp iM_W \hPi^{\pm}_{\nu}(q)& = 0 ~,\nonumber \\ 
q^{\mu}\hPi^{\pm}_{\mu}(q) \pm iM_W \hPi^{\phi}(q)& = 0 ~,\nonumber \\ 
q^{\mu}q^{\nu}\hPi^W_{\mu\nu}(q) - M^2_W \hPi^{\phi}(q)& = 0 ~, 
\label{WISESM}
\eea
while the PT vertices satisfy
\be
q^{\mu}\hGa^{W^{-}d\bar{u}}_{\mu}(q,k,-k-q) 
+ iM_W \hGa^{\phi^{-}d\bar{u} }(q,k,-k-q)
= \frac{g}{\sqrt{2}}\left[\hSi^d(k)P_L - P_R\hSi^u(k+q)\right]~,
\label{WIVESM1}
\ee
and 
\be
q^{\mu}\hGa^{W^{+}u\bar{d}}_{\mu}(q,k,-k-q) 
- iM_W \hGa^{\phi^{-}d\bar{u} }(q,k,-k-q)
= \frac{g}{\sqrt{2}}\left[\hSi^u(k)P_L - P_R\hSi^d(k+q)\right]~.
\label{WIVESM2}
\ee
Using the elementary decomposition 
\be
\Delta^{\mu\nu}_{i}(q,\xi_j)=
U^{\mu\nu}_{j}(q)-\frac{q^{\mu}q^{\nu}}{M^{2}_{j}}\Delta_{s}(q,\xi_{j})~~,
\label{Id1}
\ee
where $j=W,Z,\gamma$,  
we observe that all the remaining $\xi_t$ dependence is carried by 
the propagators of the unphysical scalars. 
Recalling the current relations: 
\be
q_{\mu}J^{\mu}_{W} = -iM_WJ_{\phi}~,~~~~~~
q_{\mu}J^{\mu\dagger}_{W} = iM_WJ^{\dagger}_{\phi} ~, 
\label{divcurrent}
\ee
it is easy to observe that by virtue of the WI of Eq.(\ref{WISESM}),
Eq.(\ref{WIVESM1})
and  Eq.(\ref{WIVESM2})
this residual  $\xi_t$ dependence cancels. Finally, as advocated, the amplitudes 
$\hat{T}_1,~\hat{T}_2$ and $\hat{T}_3$
are independent of {\it both}
$\xi_l$ and $\xi_t$.

\section{Discussion and Conclusions}

The  analysis presented in the previous sections shows    
rather transparently the mechanism responsible for the 
dual gauge cancellations of the S-matrix. 
In summary, the one-loop Feynman diagrams of an \Sm reorganize 
themselves systematically via the PT algorithm, which relies on the
full exploitation of tree-level WI. At the end of the PT algorithm  
all gauge dependences inside loops has cancelled, giving rise
to effective GFP-independent \Gfs. These one-loop effective \Gfs  
satisfy their tree-level WI,  which in turn enforce the elimination of  
all remaining gauge dependences, appearing  
outside of the loops. 
Consequently, 
one can freely choose  {\it different} 
gauge parameters $\xi_l$ and $\xi_t$,
to gauge fix the bosonic propagators appearing inside and outside 
of quantum loops, respectively.

It would be interesting to understand this dual choice of gauges 
at a more formal level; this is however beyond our power at this point.
The only known context where such a dual choice of gauge fixing parameters can 
be
field-theoretically justified is the Background Field Method (BFM) 
\cite{Abbott} .
In the BFM framework the gauge field is split into two pieces,
a ``background'' field (which corresponds to the field we call ``tree'' 
in this paper)
and a ``quantum'' field (corresponding to our ``loop'' gauge field).
It turns out that the background and quantum fields can be gauge-fixed using 
to completely
independent gauge fixing terms, which in turn, introduce two independent gauge 
fixing parameters,
$\xi_{C}$ and $\xi_{Q}$. 
 The gauge fixing procedure
is chosen in such a way as to retain the original gauge invariance 
for the background fields; consequently, $n$-point functions involving 
background fields
satisfy naive, tree-level WI to all orders in perturbation theory.  
By virtue of this last property one can show that in the BFM formulation 
the $S$-matrix is independent
of both $\xi_{C}$ and $\xi_{Q}~~\cite{BFM}$.
The analysis presented in this paper however precisely points to the fact that
the DGF property holds {\it regardless} of the gauge fixing procedure used to 
quantize 
the theory. Indeed, nowhere throughout the paper have we resorted to the 
BFM formalism. From this point of view, the DGF should be regarded as a 
general property 
of the $S$-matrix, rather then 
a property linked to some sophisticated gauge fixing procedure.
 
It is plausible that the DGF property holds true to all orders in perturbation 
theory; so far we can only show its validity at 
the one loop level since the PT has thus far been implemented only at 
one loop.
 
We believe that the PT in general, and the DGF property in particular, 
will be very useful in the implementation of automatic 
codes for calculating one-loop cross sections \cite{codes}. 
The advantages of writing one-loop amplitudes in a manifestly gauge 
independent way, as dictated by the PT, are numerous : 

(i) All UV divergences reside in the self energy functions only, while the 
improper vertices are UV finite. 

(ii) In the self energies, bosonic and fermionic contributions are 
treated in an equal footing. Furthermore, the PT self energies can be 
Dyson resumed,  giving rise to the running 
couplings of the theory \cite{Cornwall},\cite{PP}. In addition, their imaginary parts 
provide the natural regulator for resonant 
amplitudes, i.e. amplitudes containing unstable particles \cite{PP2}, \cite{PP}. 

(iii) Since each class of diagrams is rendered gauge parameter 
independent analytically, large gauge cancellations, which may 
significantly slow down 
the numerical computations, are thus avoided. A characteristic example
is the unitarity of the process $e^{+}e^{-}\rightarrow W^{+}W^{-}$.
In this case,  the contributions to the cross section of 
the electromagnetic and weak dipole moment form factors of the $W$, 
stemming from  the conventional vertex graphs,  
grow monotonically
with the momentum transfer $s$ ~\cite{TGV}; it is only after the appropriate contributions
from box diagrams have been identified by the PT and added to the vertex 
that one arrives at expressions for the form factors  which respect unitarity \cite{TGVP}. 
Even though all such pieces exist
in the $S$-matrix anyway, the advantage of carrying out the cancellations      
analytically, before resorting to numerical integrations, is obvious. 

(iv) As far as one-loop calculations are concerned, the DGF property 
results in the following simplification. For the tree bosons one 
is free to use the unitary gauge $(\xi_t\rightarrow \infty)$ while 
for the loop bosons one can use the Feynman gauge for example $(\xi_l = 1)$. 
The advantage is two-fold: 
since only physical particles 
appear in the unitary gauge, the number of diagrams is significantly reduced,  
while, at the same time, manifest renormalizability is still retained, 
because the 
loop integrals are evaluated in the Feynman gauge. 

\medskip
{\bf Acknowledgments.}~
One of us (J.~P.) thanks A.~Pilaftsis and J.~Watson for useful discussions.


\begin{thebibliography}{99}
\bibitem{GIS}
B.~W.~Lee and J.~ Zinn-Justin,
Phys.~Rev. {\bf D 7}, 1049 (1973); \\
E.~S.~Abers and B.~W.~Lee,
Phys. Rep. {\bf 9C}, 1 (1973);\\ 
B.~W.~Lee in {\it Methods in Field Theory, Les Houches 1975}, edited by
R.~Balian and J.~ Zinn-Justin (Amsterdam, North Holland); \\
C.~Becchi, A.~Rouet, and R.~Stora,
Ann. Phys, {\bf NY 98}, 287 (1976).
\bibitem{Cornwall}
J.~M.~Cornwall,                              
in {\sl Proceedings of the French-American Seminar on Theoretical
Aspects of Quantum Chromodynamics},
Marseille, France, 1981, edited J.~W.~Dash (Centre de Physique
Th\'eorique, Marseille, 1982); \\
J.~M.~Cornwall,
 Phys. Rev.; {\bf D 26}, 1453 (1982)
\bibitem{PTvarious}
J.~M.~Cornwall and  J. Papavassiliou ,
 Phys. Rev. {\bf D 40}, 3474 (1989); \\
J.~Papavassiliou,
 Phys. Rev.; {\bf D 41}, 3179 (1990); \\
G.~Degrassi and A.~Sirlin,
 Phys. Rev.; {\bf D 46}, 3104 (1992);\\
J.~Papavassiliou and C.~Parrinello,
Phys.~Rev.~D50, 3059 (1994); \\ 
J.~Papavassiliou and K.~Philippides, 
Phys.~Rev.~D52, 2355 (1995); \\
K.~Hagiwara, S.~Matsumoto, D.~Haidt, and C.~S.~Kim,
Z. Phys. {\bf C64}, 559 (1994); \\
N.~J.~Watson, \PLB{\bf 349}, 155 (1995).
\bibitem{axial} 
W.~Kummer, {\it Acta Phys. Austr.} {\bf 14}, 149 (1961);\\ 
R.~L.~Arnowitt and S.~I.~Fickler, {\it Phys. Rev.} {\bf 127}, 1821 (1962);\\
W.~Kummer, {\it Acta Phys. Austr.} {\bf 41}, 315 (1975); \\
Yu.L. Dokshitzer, D.I. Dyakonov, and S.I. Troyan, {\it Phys.\
Rep.}\ {\bf 58}, 269 (1980); \\ 
A. Andra$\breve{\mbox{s}}$i and J.C. Taylor, {\NPB}{\bf 192}, 283 (1981);\\ 
D.M. Capper and G. Leibbrandt, {\PRD}{\bf 25}, 1002
(1982).
\bibitem{FLS}
K.~Fujikawa, B.~W.~Lee, and A.~I.~Sanda, 
Phys. Rev. {\bf D6}, 2923 (1972). 
\bibitem{GW}
D.~Gross and F.~Wilczek,
Phys. Rev. {\bf D8}, 3633 (1973). 
\bibitem{PaBFM} J. Papavassiliou, Phys.~Rev.~D{\bf 51} (1995) 856. 
\bibitem{Sasaki}
K~Sasaki, \PLB{\bf 369} (1996) 117; \\  
K~Sasaki, NYU-TH-95-11-01, hep-ph/9512434, to appear in {\it Nucl.~Phys.} {\bf B}\\
M.~Passera and K.~Sasaki, 
{\it The gluon self-energy in the Coulomb and temporal axial gauges via 
the pinch technique}, hep-ph/ 9606274
\bibitem{PP2}
J.~Papavassiliou and A.~Pilaftsis
{\it Gauge invariant resummation formalism for two-point correlation functions}, hep-ph/9605385,
submitted to Phys. Rev.~D.
\bibitem{EW}
J. Papavassiliou , Phys. Rev. {\bf D 50} (1994) 5958.
\bibitem{Comment}
In the case of QCD we only have propagator-like pinch terms. In the electroweak
case we have also vertex-like pieces, because the currents are not conserved
(for details see \cite{EW}).
\bibitem{Abbott}
L.~F.~Abbott,
 Nucl. Phys. B 185 (1981) 189, and references therein. 
\bibitem{BFM}
A.~Denner, S.~Dittmaier, and G.~Weiglein,
 \PLB {\bf 333}, 420 (1994) ; Nucl. Phys. {\bf B440}, 95 (1995). 
\bibitem{codes}
E.~Murayama, I.~Watanabe and K.~Hagiwara, KEK Report 91-11 (January 1992); \\
T. Stelzer, W.F.~Long, {\it Comp. Phys. Com.} {\bf 81} (1994) 357; \\
 D. Perret-Gallix, 4th International Workshop on Software Engineering and
Artificial Intelligence for High Energy and Nuclear Physics (AIHENP95),
Pisa, Italy, 3-8 April 1995, hep-ph/9508235.  
\bibitem{PP}
J.~Papavassiliou and A.~Pilaftsis, {\it Phys. Rev. Let.}
{\bf 75} (1995) 3060;~ {\it Phys. Rev.} {\bf D53} 2128 (1996).
\bibitem{TGV}
E.N.~Argyres, G.~Katsilieris, A.B.~Lahanas, C.G.~Papadopoulos, 
and V.C.Spanos,
 \NPB{\bf391} (1993) 23.
\bibitem{TGVP}
J.~Papavassiliou and K.~Philippides 
Phys. Rev. D48, 4255 (1993). 
\end{thebibliography}
\end{document}